\documentstyle[prl,twocolumn,aps,epsf]{revtex}
\begin{document}
\include{psfig}

\title{Reply to the Comment on ``Charged impurity scattering limited 
low temperature resistivity 
of low density silicon inversion layers''}
\author{S. \ Das Sarma and E. H.\ Hwang}
\address{Department of physics, University of Maryland, College Park,
Maryland  20742-4111 } 
\date{\today}
\maketitle

In a recent Comment \cite{one} on our earlier preprint \cite{two}
proposing a theoretical explanation for the observed strong temperature 
dependence of the low temperature resistivity of low density ``metallic'' 
Si inversion layers \cite{three} Kravchenko {\it et al.} have raised a number 
of issues which require careful consideration. In this Reply to their 
Comment we discuss the issues raised in ref. \onlinecite{one} with 
respect to our work \cite{two} and respond to the specific questions 
about our model \cite{two} brought up by Kravchenko {\it et al.} \cite{one}.

It was conceded in ref. \onlinecite{one} that our theoretical results 
\cite{two} ``are qualitatively similar to those observed experimentally'' 
but questions were raised about our choice of parameters for the charged 
impurity density ($N_i$) and the free carrier density ($n_s -n_c$) with 
the implication that our choice of $N_i$ and $n_s - n_c$ may be 
inconsistent with the experimentally used samples in ref. 
\onlinecite{three}. We first point out that neither quantity 
($N_i$ or $n_s - n_c$) has been experimentally measured, and therefore 
the authors of ref. \onlinecite{one} are expressing their {\it opinions}
based purely on {\it  
theoretical speculations} rather than making statements of facts in ref.
\onlinecite{one}. (This distinction was not clearly made in ref. 
\onlinecite{one}.) Their remark \cite{one} on our choice of the 
charged impurity density $N_i$ is based on a naive misunderstanding 
of our theory whereas our choice of $n_s-n_c$ as the free carrier density 
(in fact, this is the definition of the effective free carrier density in 
our model \cite{two}) is based on more subtle arguments than indicated by 
the discussion of ref. \onlinecite{one}.

As emphasized in our original paper \cite{two} the parameter $N_i$ 
{\it only} sets the overall resistivity scale in our theory in the sense 
that the resistivity $\rho \propto N_i$ --- $N_i$ does not affect the 
calculated temperature dependence at all. The actual value of $N_i$ is 
set in our work by demanding overall {\it quantitative} agreement between 
theory and experiment, and the very low value of $N_i$ ($\approx 0.3 
\times 10^{10} cm ^{-2}$) reflects the anomalously high peak mobility 
($\sim 71,000 cm^2/Vs$ at $100 mK$) of the Si-15 sample, which has a 
peak mobility roughly a factor of five to seven higher than that in 
other typical good quality Si MOSFETs ($\sim 10,000 - 20,000 cm^2/Vs$). 
An independent confirmation of the low value of $N_i$ in the Si-15 sample 
comes from the empirical formula connecting $N_i$ and the peak mobility 
($\mu_m$) derived by Cham and Stern \cite{four} : 
$\mu_m = 1250 (N_i/10^{12})^{-0.79}$ where $\mu_m$ is expressed in 
$cm^2/Vs$ and $N_i$ in units of $10^{12}cm^{-2}$. This empirical formula 
leads to an $N_i \approx 0.5 \times 10^{10}cm^{-2}$ within a factor of 
two of the value $N_i \approx 0.3 \times 10^{10} cm^{-2}$ used in our 
work \cite{two}. The actual discrepancy is even less because the empirical 
formula \cite{four} refers to the $T=0$ mobility, which for Si-15 should 
be considerably higher than $71,000 cm^2/Vs$ because of the strong 
temperature dependence of the mobility observed in ref. \onlinecite{three}. 
Thus the low value of $N_i$ used in our work \cite{two} is necessitated by 
the anomalously high (almost an order of magnitude higher than usual) 
mobility of the Si-15 sample. The fact that a low value of $N_i$ is 
needed to obtain quantitative numerical agreement between our theory and 
the experimental Si-15 data \cite{three} at {\it high densities} 
($n_s \gg n_c$), 
where the standard Drude-Boltzmann transport theory should be eminently 
valid, shows rather decisively the correctness of our choice of $N_i$ in 
ref. \onlinecite{two}. We emphasize that the low value of $N_i$ used in ref.
\onlinecite{two} simply reflects the anomalously high quality of the Si-15
sample of ref. \onlinecite{three}, and nothing else.

Having established that our choice of $N_i$ in ref. \onlinecite{two} is 
quite reasonable we should now mention   two aspects of our model 
\cite{two} which seem to have been overlooked in the Comment \cite{one}.
 First, characterizing the charged impurity scattering by a single 
two-dimensional (2D) charge density of $N_i$ with all the impurity centers 
located randomly in a plane placed precisely at the Si-SiO$_2$ interface, 
as we do in ref. \onlinecite{two}, is surely a highly simplified zeroth 
order model for a complicated situation where the charged impurity centers 
will be distributed over some distance inside the oxide layer. We use the 
simple model to keep the number of free parameters to a minimum (just one, 
$N_i$). If we modify the model slightly, for example, displace the random 
charged impurity plane some finite distance inside the oxide (or consider 
a three-dimensional impurity distribution, as is likely), we could 
considerably increase $N_i$, making it sound ``more reasonable''. 
We believe that such fine tuning is unwarranted in a zeroth order model 
and should be left for future improvements of the theory. Second, our 
estimated $N_i$ is surely a theoretical {\it lower bound} on the possible 
value of the charged impurity density because our theory \cite{two} 
clearly overestimates the impurity scattering strength as we neglect 
the modification in scattering due to electron binding to the impurities, 
and assume that the charged impurities scatter via the screened Coulomb 
interaction without taking into account the binding effect discussed 
qualitatively in ref. \onlinecite{five}. 
Thus, the parameter $N_i$ in our model \cite{two} should be taken as an 
effective (single) parameter which characterizes the overall impurity 
scattering strength in the system, which should not necessarily be 
precisely the same as the number density of bare Coulomb scatterers in 
the sample.

Now , we discuss the second point raised in ref. \onlinecite{one} 
regarding our choice of $(n_s-n_c) \equiv n_e$ as the ``effective'' 
free carrier density participating in the ``metallic'' Drude-Boltzmann 
transport for $n_s > n_c$. This is a rather subtle issue because in 
our model \cite{two} $n_e$ is, {\it by definition}, the free carrier 
density entering the conductivity $\sigma = n_e e \mu$ where $\mu$ is 
the carrier mobility --- Drude-Boltzmann theory allows for an intuitive 
separation of the conductivity into an effective carrier density ($n_e$) 
and an effective carrier 
mobility ($\mu$). Our choice of $n_e \equiv n_s - n_c$ as the 
effective free carrier density leads to excellent qualitative (actually, 
semi-quantitative) agreement between our calculated theoretical results 
and {\it all} the published experimental results in Si MOSFETs and p-type 
GaAs heterostructures as far as the temperature dependent resistivity on 
the ``metallic'' side ($n_s > n_c$) is concerned. 
Our theory \cite{two}  also produces the experimentally observed 
non-monotonic temperature dependence in the resistivity arising from the 
quantum-classical crossover behavior which, we believe, to be playing 
an important role. We note that our choice of $n_e 
\equiv n_s -n_c$ as the effective free carrier density is not only 
consistent with the impurity binding/freeze-out scenario \cite{two,five}, 
but also with the percolation model of ref. \onlinecite{six}.

The important open issue is, of course, a direct experimental 
measurement of $n_e$, the effective free carrier density {\it near 
threshold} (i.e., $n_s \geq n_c$) at zero magnetic field. It has been 
known for long time \cite{seven} in MOSFET physics that a direct 
determination of the free carrier density near threshold at low 
temperatures is almost impossibly difficult (particularly if $n_s 
\leq 10^{11} cm^{-2}$ as it is in the cases of our interest), requiring 
a number of simultaneous complex measurements which, to the best of our 
knowledge, have never been attempted in the samples showing the so-called 
2D metal-insulator-transition (M-I-T) 
phenomena. Since there seems to be confusion 
or misunderstanding on this point \cite{one} we take the liberty of 
quoting from the authoritative review article \cite{eight} by 
Ando, Fowler, and Stern: ``If trapping, band tailing, or depletion charge 
are important (as they are near threshold) no single experiment can 
unambiguously give the mobility and carrier concentrations. ... . The 
interpretation of the measurements is increasingly unreliable at carrier 
concentrations below $10^{12} cm^{-2}$.'' (p. 490 in ref. \onlinecite{eight}.) 
We believe that a direct measurement of the {\it zero-field}, 
low-temperature, low-density, near-threshold free carrier density 
through careful capacitance studies (which are extremely problematic 
at low densities and 
low temperatures) is the only way to decisively settle the question 
of the effective free carrier density participating in the ``metallic'' 
transport in the Si samples of ref. \onlinecite{three}.

Following the procedure used in ref. \onlinecite{nine} for GaAs systems, 
we can however determine the operational value of $n_e$ participating in 
the Si-15 sample. We show in Fig. 1 

\begin{figure}
\epsfysize=9.cm
\epsffile{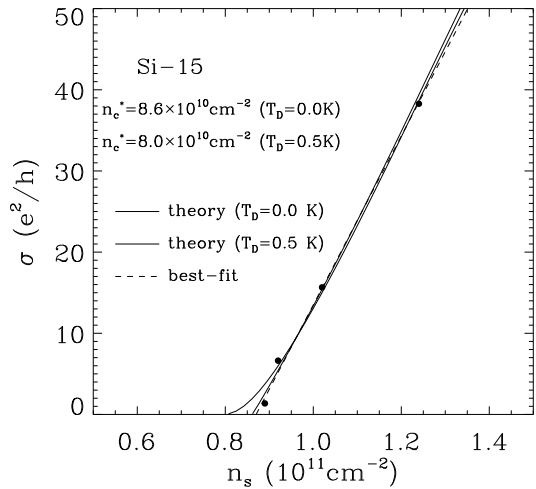}
\caption{
The calculated zero-temperature conductivity $\sigma(T=0)$
for the Si-15 sample of ref. [3] as a function of density $n_s$
using the theory of ref. [2]. The dots are the extrapolated $T=0$
experimental conductivities at respective $n_s$ values taken from ref. [3].
The dashed line is the best fit to the experimental results whereas the 
thick (thin) lines are the theoretical results using a Dingle temperature 
$T_D = 0.0 K$ ($0.5 K$) in the quantum screening (see ref. [2] for details).}
\end{figure}

\noindent
the extrapolated zero temperature 
conductivity $\sigma(T=0)$ of the Si-15 sample\cite{three} plotted as 
a function of the carrier density $n_s$, both for experimental 
data\cite{three} and our theoretical results \cite{two}. It is clear 
that $\sigma(T=0) \propto (n_s - n_c^*)$ in Fig. 1 where $n_c^* \approx 
n_c$. We believe that this specific functional behavior of the $T=0$ 
conductivity, vanishing approximately linearly in $(n_s-n_c)=n_e$ near 
threshold, 
provides compellingly strong phenomenological support 
for our model of 
using $n_e=n_s-n_c$ as the {\it effective} free carrier density in the 
Drude-Boltzmann theory. Note that in 
our model $\sigma = n_e e \mu$ with $n_e=n_s-n_c$, and the calculated 
$\mu$ has
very weak $n_s$ dependence near threshold (see Fig. 1) --- this is strong 
support for our assumption that $n_e = n_s -n_c$ provides a measure of 
the effective free carrier density in the theory.
We have found \cite{ten} the same behavior in {\it all} the published Si data.

Third, we address the issue of the unpublished {\it finite magnetic field} 
(e.g. Hall effect) measurement of the free carrier density alluded to in 
ref. \onlinecite{one}. This connection made in ref. \onlinecite{one} is, 
in fact, misleading since the application of an external magnetic field 
to a 2D system {\it completely} changes its physics by converting it to 
a quantum Hall system whose localization properties are still poorly 
understood. The free carrier density ($n_e$) entering our Drude-Boltzmann 
theory\cite{two} is, by definition, the {\it zero-field} free carrier density 
which is not necessarily the same as that measured in {\it finite field} quantum 
Hall-type experiments alluded to in ref. \onlinecite{one}. {\it This fact 
is most easily demonstrated by pointing out that the finite field 
measurements of the type mentioned in ref. \onlinecite{one} give the free 
carrier density to be $n_s$ even deep inside the localized regime ($n_s < n_c$) 
where the free carrier density in the Drude-Boltzmann theory is 
obviously zero, $n_e = 0$ in the insulating phase.} Thus the finite-field 
carrier density measurements give the free carrier density to be $n_s$ 
{\it everywhere} (both in the metallic and the insulating phase), and 
cannot therefore have anything to do with the 
parameter $n_e$ entering our Drude-Boltzmann theory. While 
understanding these interesting 
finite field measurements is clearly an important 
theoretical challenge (well beyond the scope of our manifestly zero field 
theory \cite{two} aimed exclusively at understanding the temperature 
dependent resistivity on the ``metallic'' side), we disagree with the 
suggestion in ref. \onlinecite{one} that an understanding of the 
zero-field transport properties, as in ref. \onlinecite{three}, must 
somehow depend crucially on a complete theoretical understanding of 
the quantum Hall behavior of these systems.
(None of the proposed theoretical models for 2D M-I-T can account 
for the finite field quantum Hall behavior.)

Finally, we mention that characterization of our theory as a ``simple 
classical model'' in the concluding sentence of ref. \onlinecite{one} 
is misleading. While quantum interference 
corrections are neglected in our theory (we argue that quantum 
interference corrections 
are overwhelmed by the screening effects considered in our theory in the 
$T \geq 100 mK$ temperature range considered in the experiments 
\cite{three}), the striking temperature dependence \cite{two} of our 
theoretical results arises entirely from the classical-quantum crossover 
phenomenon 
and the quantum screening of charged impurity scattering -- a classical 
theory would not have any of the effects obtained in our calculation 
\cite{two}.

This work is supported by the U.S.-ARO and the U.S.-ONR.

\end{document}